\documentclass{osa-article}
\journal{oe}
\articletype{Research Article}
\usepackage{lineno}
\usepackage[utf8]{inputenc}
\usepackage{subcaption}
\usepackage{bm}
\usepackage{mathtools}
\usepackage{xcolor}
\usepackage{tabularx}
\usepackage{algorithm}
\usepackage{algpseudocode}
\usepackage{booktabs}

\newcommand{\D}{\mathrm{d}}
\DeclareMathOperator*{\argmin}{arg\,min}

\usepackage[hang,flushmargin]{footmisc}
\usepackage{lipsum}
\makeatletter
\newcommand{\algorithmfootnote}[2][\footnotesize]{%
  \let\old@algocf@finish\@algocf@finish
  \def\@algocf@finish{\old@algocf@finish
    \leavevmode\rlap{\begin{minipage}{\linewidth}
    #1#2
    \end{minipage}}%
  }%
}
\makeatother

\begin{document}

\title{Matched Illumination}
\author{Yuteng Zhu,\authormark{1,2} and Graham D. Finlayson,\authormark{1,3}}
\address{\authormark{1}Colour and Imaging Laboratory, University of East Anglia, Norwich Research Park, Norwich, NR4 7TJ, United Kingdom\\
\authormark{2}yuteng.zhu@uea.ac.uk\\
\authormark{3}g.finlayson@uea.ac.uk\\}


\begin{abstract}
In previous work, it was shown that a camera can theoretically be made more colorimetric - its RGBs become more linearly related to XYZ tristimuli - by placing a specially designed color filter in the optical path. While the prior art demonstrated the principle, the optimal color-correction filters were not actually manufactured.
In this paper, we provide a novel way of creating the color filtering effect without making a physical filter: we modulate the spectrum of the light source by using a spectrally tunable lighting system to recast the prefiltering effect from a lighting perspective.  According to our method,  if we wish to measure color under a D65 light, we relight the scene with a modulated  D65 spectrum where the light modulation mimics the effect of color prefiltering in the prior art.   We call our optimally modulated light, the \emph{matched illumination}.  
In the experiments, using synthetic and real measurements,  we show that color measurement errors can be reduced by about 50\% or more on simulated data and 25\% or more on real images when the matched illumination is used.
\end{abstract}

\section{Introduction}
Digital cameras measure the color information in a real-world scene {\it like} a human observer only if a camera meets  the  Luther condition~\cite{ives1915,luther1927}. The Luther condition requires the camera sensitivity functions are a linear combination of the color matching functions of the human visual system~\cite{HORN1984}. If a camera meets the Luther condition, the colors it measures are linearly related to the device-independent tristimulus values, such as CIE XYZ tristimuli \cite{ohta2006colorimetry}. Such a camera is said to be colorimetric. However, typical cameras do not satisfy the Luther condition and so cannot be used for precise color measurement~\cite{nakamura2016image}.  

A viable way to improve the colorimetric accuracy of a camera is to capture multiple images, each with a different color filter placed in front of the camera~\cite{farrell1995method,wu2000imaging}. This multi-shot technique can gather more color information  than in a single shot (greater than 3-dimensional color signals) and when they are mapped to colorimetric XYZ values, we can obtain greater accuracy. Generally, the color filters are chosen from commercial products~\cite{macadam1945colorimetric,hardeberg2004filter,Imai2001DigitalCF} either heuristically or by using an exhaustive search process~\cite{vora1997design}. An alternative way is to capture images under multiple lights, such as using a light booth with different illuminants \cite{xu2016filter,martinez2019spectral}.
Here, the multiple lights perform an analogous role as the filters. However, both methods require multiple shots of images which take a longer capture process and if nothing else, the registration between images is a problem itself. 

\begin{figure}
    \centering
    \begin{subfigure}{\textwidth}
    \centering
    \includegraphics[width=\textwidth]{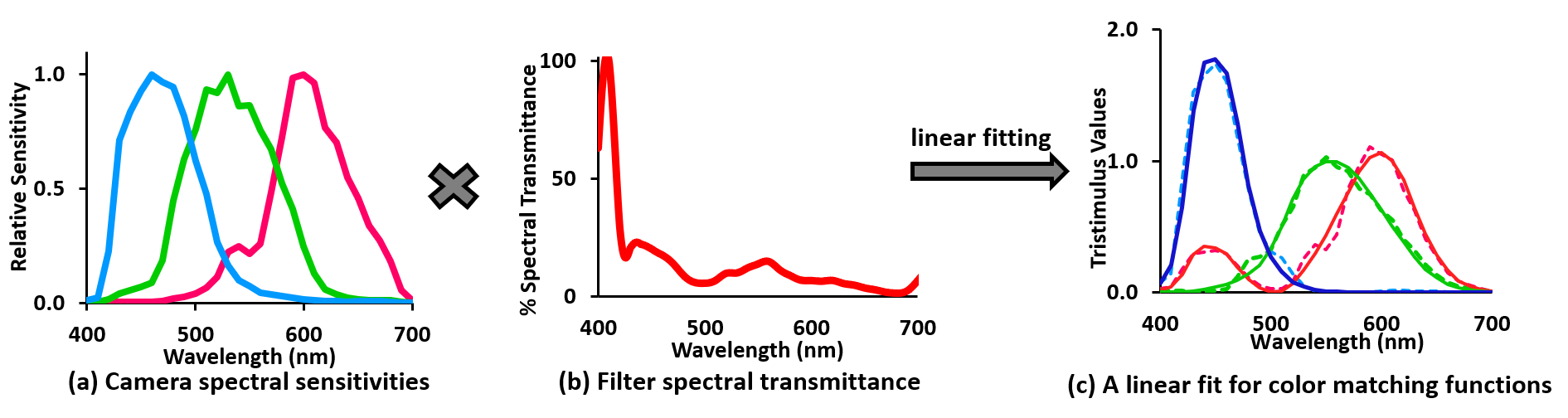}
    \caption*{top panel: the Luther condition of a given camera is improved by a color filter}
        \label{fig:fig1-1}
    \end{subfigure} \hfill
	\hfill
\begin{subfigure}{\textwidth}
    \centering
    \includegraphics[width=\textwidth]{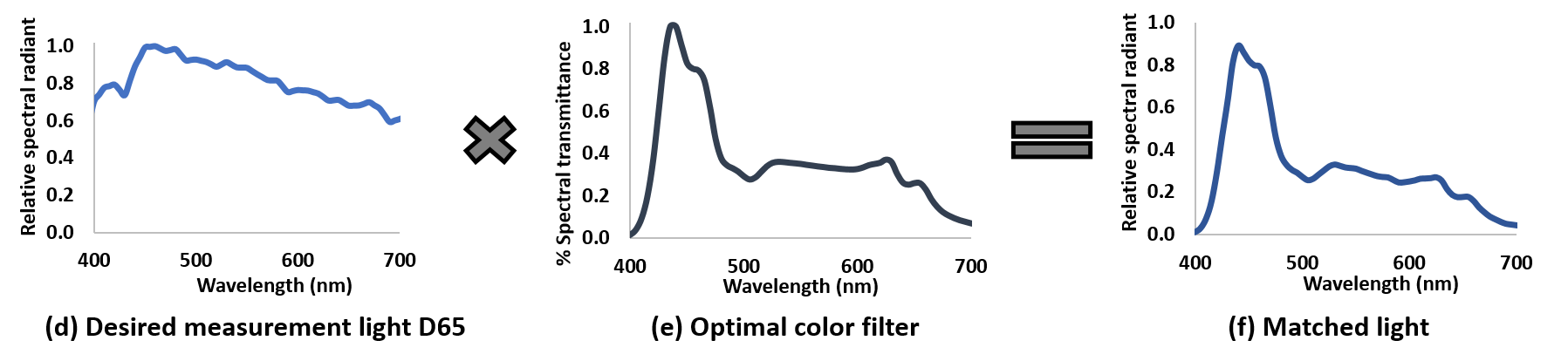}
    \caption*{bottom panel: a matched D65 illumination that modulates from D65 by a color filter}
        \label{fig:fig1-2}
    \end{subfigure} \hfill
    \caption{In the top panel, we show the filter-modified Luther-condition. Given a camera with known RGB sensitivities as in (a), an optimal filter (b) can be found that after a linear regression fit, the corrected camera sensitivities (dashed lines in (c)) are good approximation to the XYZ color matching functions (solid lines in (c)). A matched illumination (f) is determined given the spectral characteristics of the desired measurement light (d) and the optimal color filter (e).}
    \label{fig:fig1}
\end{figure}

Finlayson and Zhu \cite{TIP2020filter,filter2020sensor} recently proposed to improve the colorimetric accuracy of a digital camera by placing a carefully-designed color filter in the optical path way with a single-shot image. The spectral transmittances of such a filter can be optimally designed for making a camera better meet the Luther condition. In Figure~\ref{fig:fig1}, the top row illustrates how a  color-correction filter can make a camera more colorimetric. Figure 1a shows the spectral sensitivities of a camera, which are notably different from the CIE XYZ color matching functions (see the solid lines in Figure 1c). The physical effect of placing a color filter - such as the one shown in Figure 1b -  in front of a camera can be reasonably modeled as the multiplication of the filter spectral transmittance and the camera sensitivities on a per wavelength basis over the visible spectrum. After prefiltering, the camera sensitivity functions are linearly fitted to the reference CIE XYZ color matching functions \cite{ohta2006colorimetry}. The corrected sensitivities are shown in Figure 1c. The solid lines show the reference color matching functions and the dashed lines show the effective camera sensitivities after the best linear fitting. Clearly, we see that by using a color filter we can make the camera curves a close approximation to the visual sensitivities. At the time of writing this manuscript, it is not known whether the optimal color filters can, in fact, be manufactured.

Our contribution begins with the observation that for typical color measurement scenarios, the effect of a color filter placed in front of the camera can be achieved by placing the same filter in front of the light source. We call the modulated light source the \emph{Matched Illumination}. 
It follows that if we wish to measure colors under a target measurement light source, say the standard daylight of D65 (see the bottom panel in Figure \ref{fig:fig1}), we need to match it to a new illumination, effectively a filtered D65 (that is not D65). Then the camera will capture the object colors using this matched illumination to predict the ground-truth XYZ tristimuli of the desired measurement light source. Note that the derived filter shown in panel 1e is not the same as 1b since it is derived with respect to a tunable LED illuminator (discussed below). This illuminator places more physical constraints on the design of the filter and matched illumination compared with the original camera+filter work.

In this paper, our theory of matched illumination  is developed using a Gamma Scientific RS-5B spectral illuminator. The Gamma Scientific illuminator has eight narrow and two broad band LEDs. We will show how, for a given light (produced by the illuminator), we can solve for the best matched illumination. While our algorithm development is tied to the Gamma Scientific illuminator, the techniques are generally phrased and so could be deployed to other multi-band lights. As an important detail, we deal with the ornery issue that the spectral shape of LED outputs shift as the driving voltage changes. 

The work of \cite{Wang2021} is related to our approach. There, a similar illuminator is tuned - by means of a genetic algorithm -  to find a spectrum of light that better allows reflectance spectra to be recovered from camera RGBs.
Though,  for the SFU reflectance set \cite{barnard2002data} (used in \cite{Wang2021} and our study), the spectra recovered under their derived optimal illuminant are no more accurate than the spectra recovered under a fixed illuminant. 

Experiments validate our approach. We show that we can significantly reduce color measurement error for a desired measurement light by solving for and then measuring with respect to the matched illumination. A novel aspect of our experimental work is that we develop and deploy a novel new technique for generating large spectra data sets given only a small number of spectral measurements. We exploit the idea that - in raw image capture - the RGBs computed from a linear combination of RGBs - up to noise - must be the same as the single RGB measured by viewing a linear combination of the underlying reflectances. Using this idea, we generate the RGBs for the large set of 1995 reflectances (SFU reflectance data set) using only 24 RGBs measured in a Macbeth ColorChecker chart \cite{mccamy1976chart}. 

In Section \ref{sec:background}, we present the prior art to our method as well as the relevant background on image formation. In Section  \ref{sec:method}, we present our method for calculating the matched illumination. Experiments are reported in Section \ref{sec:exp}. In Section \ref{sec:conclusion}, there is a short conclusion.

\section{Background} \label{sec:background}

\subsection{Color Formation with a Filter}
The physical process of forming a color pixel underpins our idea of illumination matching. The color recorded by a digital camera mainly depends on  the light stimulus, the object reflectance, and the sensitivity responses of the camera. They are respectively  represented by the spectral functions $ E(\lambda)$, $R(\lambda)$, and $Q_{k}(\lambda)$. The RGB response is written as:
\begin{equation}
    \label{eq:pixelformation}
	\rho_{k} = \int_{\omega} \! R(\lambda) E(\lambda) Q_{k}(\lambda)\,\D\lambda \,, \quad k \in \{R, G, B\}
\end{equation}
where $\rho_{k}$ denotes one of the RGB color values. Here and henceforth, $\lambda$ denotes the wavelength variable defined over the visible spectrum  $\omega$.

When a transmissive color filter $F(\lambda)$ is placed in the optical pathway, the filtered RGB is written as:
\begin{equation}
    \label{eq:filtered_pixel}
	\rho^{filtered}_{k} = \int_{\omega} \! R(\lambda) E(\lambda) F(\lambda)Q_{k}(\lambda)\,\D\lambda \,, \quad k \in \{R, G, B\}
\end{equation}
where $F(\lambda)$ denotes the spectral transmittance of the filter with respect to the wavelength variable.

It is useful to sample spectral data and describe them in the discrete vector-matrix representation. Let $\mathbf{Q}$ denote the spectral sensitivities of a camera.
The columns in the matrix represent the spectral sensitivity functions for each sensor channel and the rows denote the sensor responses at sampled wavelengths. Hence,  $\mathbf{Q}$ is an $n \times 3$ matrix where $n$ is the number of sampled points across the visible range. In this paper, the spectral data are collected in the visible range from 400\,nm to 700\,nm for every 10\,nm. Thus, we have $n=31$.

Similarly, let the 31-vectors $\mathbf{e}$ and $\mathbf{r}$ denote sampled representations of a light and a surface. Let $diag()$ denote the function which takes an $n$-vector as an argument and maps it to an $n \times n$ diagonal matrix. We can rewrite the image formation in Equation (\ref{eq:pixelformation}) as:
\begin{equation}
\boldsymbol{\rho}=\mathbf{Q}^T diag(\mathbf{e})\mathbf{r}
\label{eq:pixelformation1}
\end{equation}
where we assume the wavelength sampling is incorporated in $\mathbf{Q}$ and $\boldsymbol{\rho}$ is a $3 \times 1$ vector denoting the RGB triplet values.

\subsection{The Luther Condition}

A camera is said to be colorimetric if it satisfies the Luther condition: the camera sensitivities are a linear combination of the standard color matching functions~\cite{HORN1984}. Let $\mathbf{X}$ denote the $31\times 3$ matrix where the columns are the X, Y and Z color matching functions (again we sample from 400\,nm to 700\,nm at a 10\,nm sampling interval). In this discrete representation, the Luther condition is written as:
\begin{equation}
    \mathbf{X = QM} 
    \label{eq:Luther}
\end{equation}
where $\mathbf{M}$ is a $3 \times 3$ full rank matrix denoting the linear transform between two  sets of sensitivities.

The Luther condition is rarely met by  an off-the-shelf digital camera. In \cite{TIP2020filter}, we proposed a new filtered version of the Luther condition. If there exists a color filter vector $\mathbf{f}$ such that:
\begin{equation}
    \mathbf{X} = diag(\mathbf{f})\mathbf{QM} 
    \label{eq:filtered-Luther}
\end{equation}
then the {\it Filtered Luther condition} is met. 

Of course neither the Luther condition nor the filtered variant is likely to hold exactly. Thus, a key focus of the prior art work on filter design \cite{TIP2020filter} was to develop the numerical methods to find filters that make cameras most colorimetric, i.e.\ that make them best satisfy the Luther condition.

\subsection{Color Correction}

To use a camera for color measurement - whether we use a color filter or not - the recorded camera RGBs are color corrected to XYZ counterparts using a $3\times 3$ correction matrix. While other non-linear color correction methods could be used (e.g.\ \cite{hong2001study,finlayson2015color,hung1993colorimetric,andersen2016weighted}), a linear color correction has several advantages. First, based on arguments from image formation, a $3\times 3$ matrix correction should work well \cite{drew1992natural}. Second, a linear transform is scalar invariant. If we double the illumination intensity that lights a scene, then the corresponding RGBs and XYZs also double and the goodness of fit afforded by a $3\times 3$ matrix remains unchanged. Finally, if colors fall on a line in the RGB space, they still fall on a line after color correction (an important physical consideration for correctly mapping highlights in photographic images \cite{andersen2016weighted}).

\subsection{The Gamma Scientific RS-5B Illuminator}

\begin{figure}
    \centering
    \includegraphics[width=0.9\textwidth]{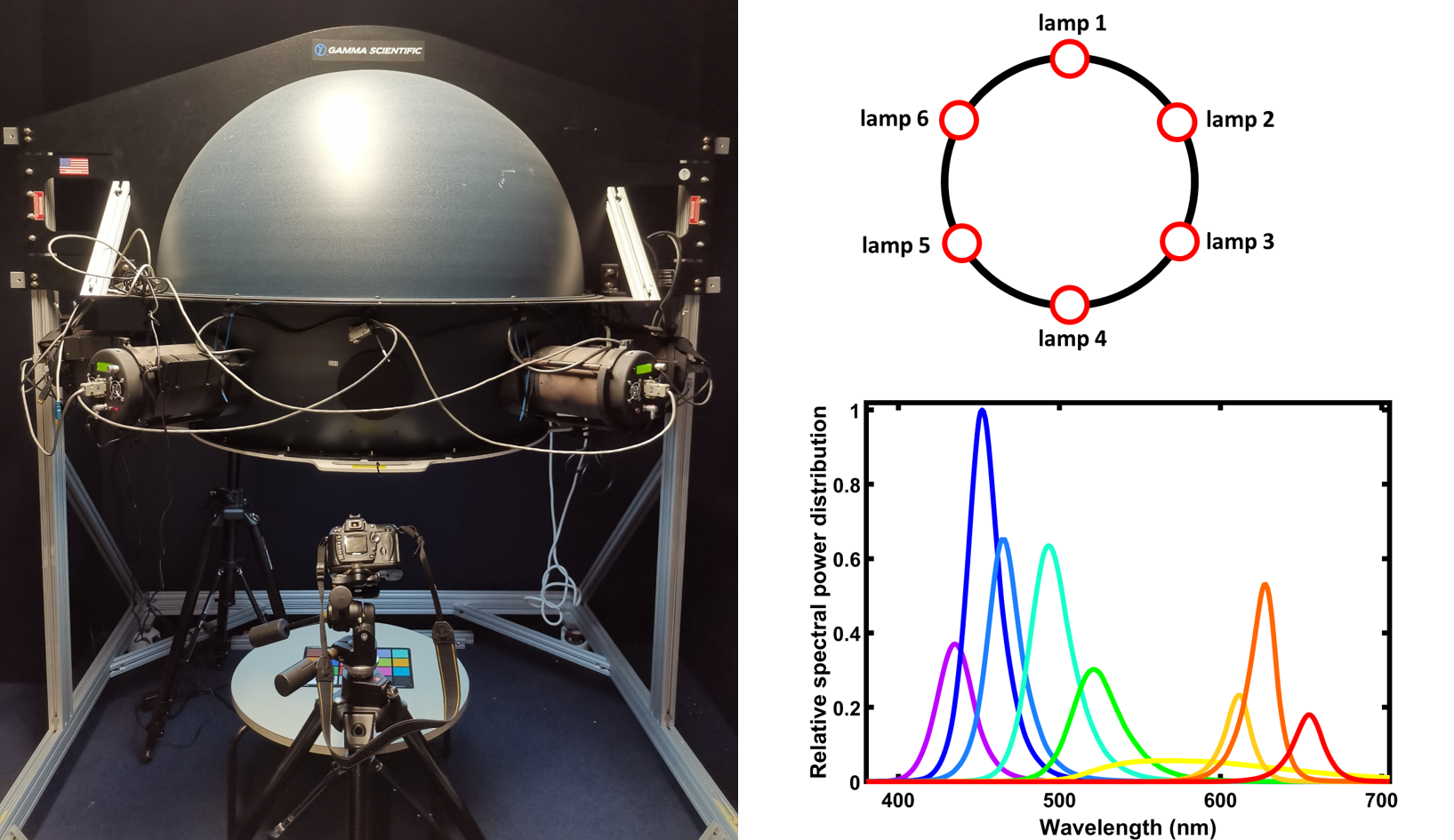}
    \caption{On the left, we show the experimental setup: a digital camera is set on a tripod to capture the image of the object on the table illuminated by a desired light generated by the illuminator system. On the back of the half-sphere illuminator, a tele-spectroradiometer is used to measure the spectrum of the light. The illuminator consists of six lamps arranged in the integrating sphere. Its sectional arrangement is drawn on the top right.  Each lamp has 10 LED channels and their relative spectral power distributions at their maximum intensity are plotted on the bottom right.}
    \label{fig:fig2}
\end{figure} 
    
The Gamma Scientific illuminator system has six lamps carefully arranged in the perimeter of the integrating sphere with white diffusing boards installed inside for creating spatially uniform lighting, see Figure~\ref{fig:fig2}. Each lamp consists of 10 different LED channels. The spectral power distributions of each LED channel (when the maximum current is driven) are shown in the bottom right of Figure~\ref{fig:fig2}. Note that only nine spectra can be seen in the figure as two broad LED lights have almost the same spectral power.  From the figure, we can see that eight of them are narrow-band LED lights ranging from  blue to red while two are identical yellowish broad-band LEDs. A broadband LED is used because of the lack of green LEDs in the range between 525\,nm to 615\,nm.

The intensity of each LED light can be digitally controlled and programmed (using a serial communication port) in any combination and proportion to generate a desired illumination spectrum. Ideally, the light spectrum driven at partial intensity should have the same spectral shape only with a scaling factor as that driven by the maximum intensity. In such a condition, we say the spectrum scales linearly with the intensity levels. When the linearity holds and the light spectra at its maximum intensity are measured, we are able to predict the illumination spectrum when we program the intensity levels of the light sources. 

\begin{figure}
    \centering
    \begin{subfigure}{0.45\textwidth}
    \centering
    \includegraphics[width=\textwidth]{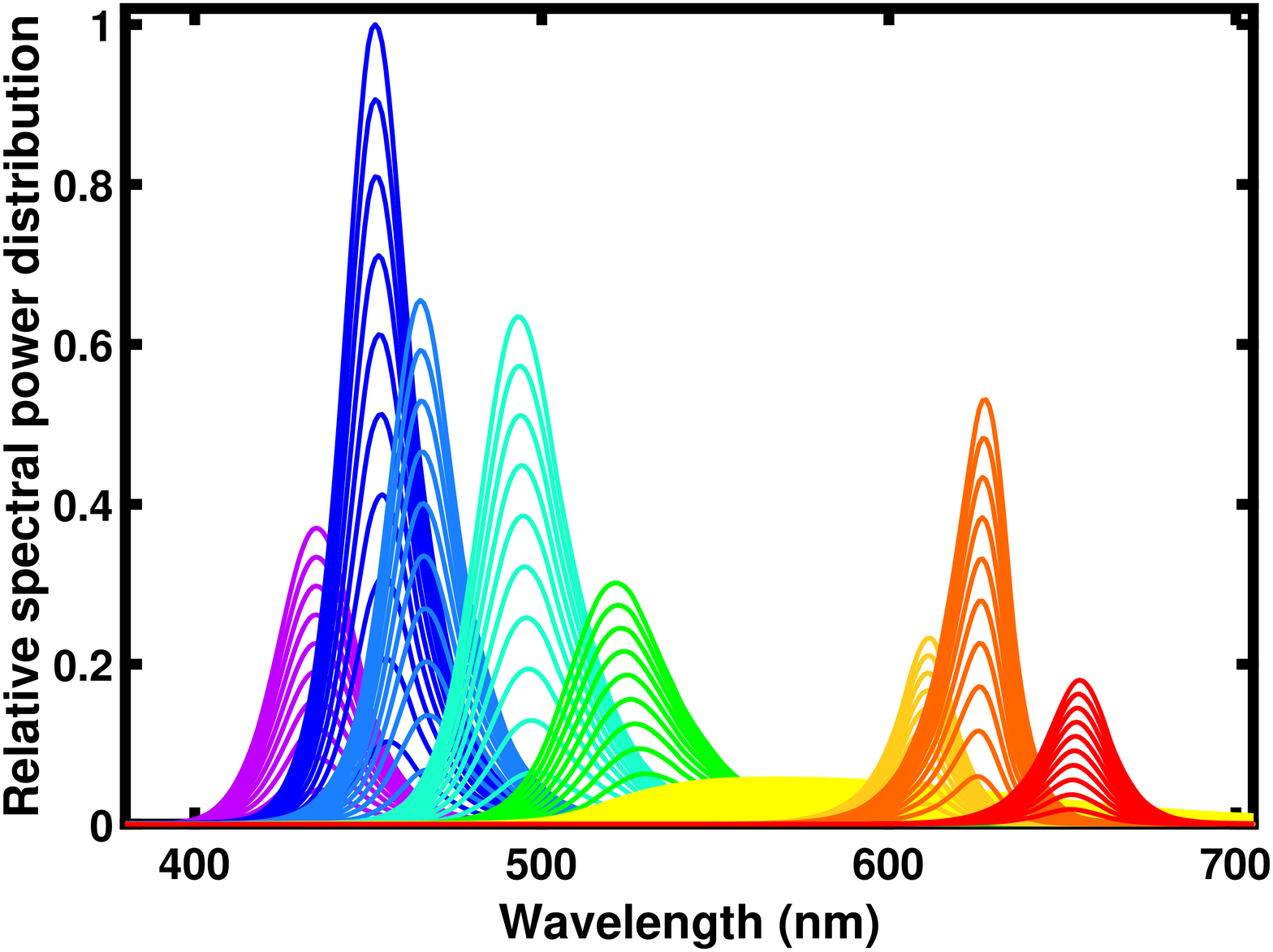}
    \caption{Spectral distributions at varied intensities}
        \label{fig:fig3-1}
    \end{subfigure} 
	\hfill
\begin{subfigure}{0.45\textwidth}
    \centering
    \includegraphics[width=\textwidth]{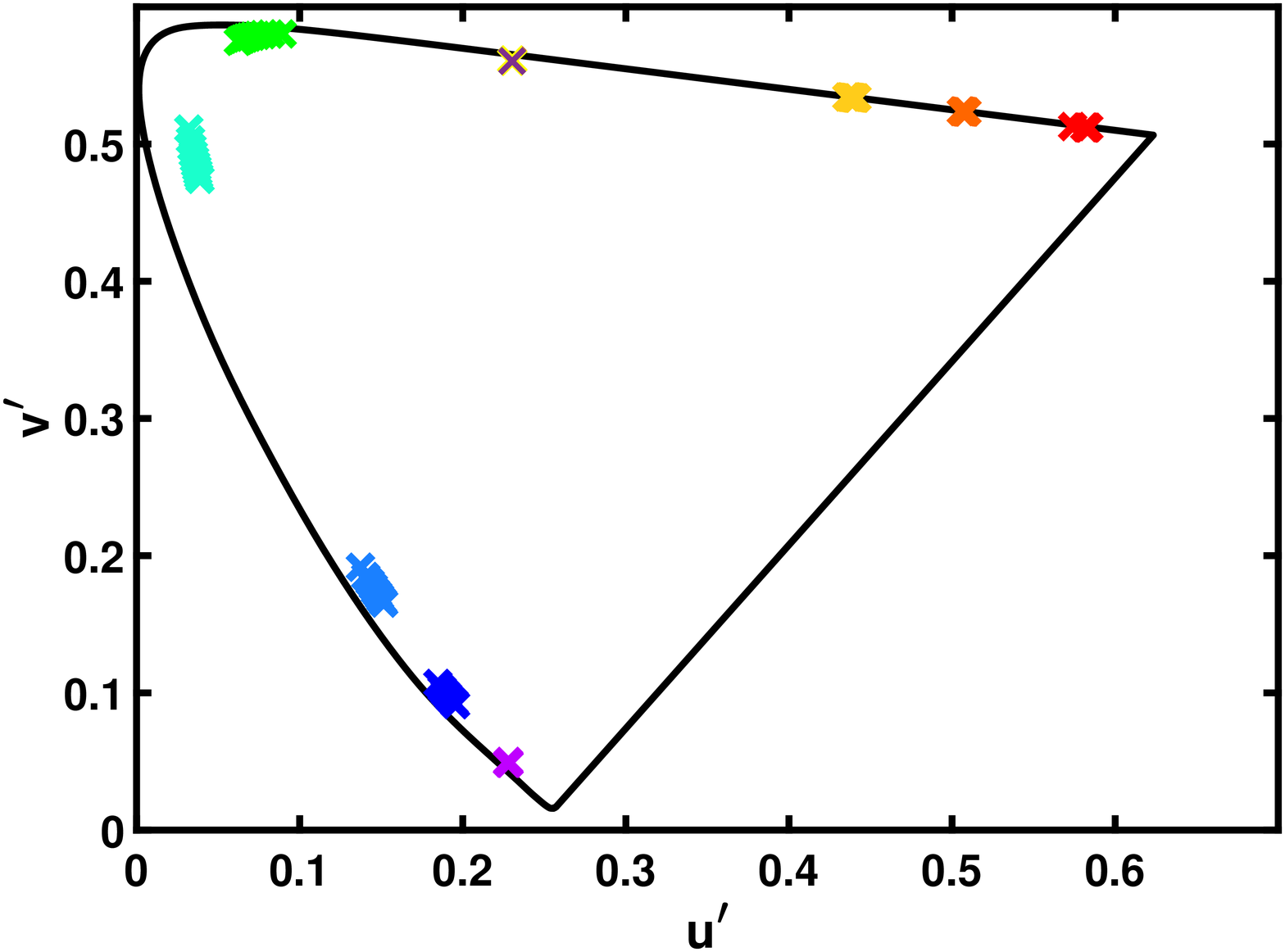}
    \caption{Chromaticity coordinates  at varied intensities }
        \label{fig:fig3-2}
    \end{subfigure} 
    \caption{The relative spectral power distributions at varied intensity levels are plotted in (a). Their u'v' coordinates are plotted in the chromaticity diagram in (b). Note the horse-shoe shaped outline in (b) is the spectral locus.}
    \label{fig:fig3}
\end{figure}

However, in practice, when we adjust the intensity level (driving current) of the light sources, we find that, for some LEDs, the peak wavelength of the spectra shifts. So, we characterize the illuminator system by measuring the spectral distributions of each light source at varied driving current levels between 0\% and 100\% of its full intensity, i.e.\ $[0, 0.1, 0.2, \cdots, 1]$. Their spectral distributions are plotted in Figure~\ref{fig:fig3-1}. It can be seen that there is some shift in the peak wavelength when intensity level changes. For example, as intensity decreases, the peaks of the fifth  (from left to right in Figure~\ref{fig:fig3-1}) LED channel shown in green lines slightly shift towards the longer wavelength. The shift reaches 17\,nm between the maximum and minimum intensities.
We also calculate the u'v' chromaticity coordinates \cite{hunt2011measuring} for all intensity levels for each LED channel and plot in the chromaticity diagram, see Figure \ref{fig:fig3-2}. Each LED channel is depicted by one color. We can see 9 colored clusters with respect to 9 LED types. Among them, we see that two LEDs in the green-cyan area have noticeable chromatic shift while others are relatively stable (e.g.\ red LEDs).  When the chromatic shift is significant, we can no longer predict the illumination spectrum under the assumption of linearity.

As a final comment, returning to Figure \ref{fig:fig2}, we see that different LEDs have significantly different power ranges. The importance of this physical feature is that it places a constraint on the spectral power distribution of any matched illumination. Indeed, for us to replicate the prior art work on transmissive filters in the lighting world, we would need narrow band lights across the visible spectrum that had the same peak maximum intensities. Thus, a priori we expect our matched lights to perform less well than unconstrained optimized filters. This said, our matched lights have the advantage over the prior filter design work that they can be - as we show next - physically realized.

\subsection{Optimized Illumination}

Before presenting our method, we wish to point the reader to  prior art reported in the literature. 
In \cite{Wang2021}, a lighting system with spectrally-tunable LEDs was used for the spectral reconstruction (SR) problem. In SR, we attempt to recover spectra from camera RGB responses. In \cite{Wang2021}, the best composition of the LED lights was sought that subserves the SR task.  For a variety of different regression-based SR algorithms, a genetic algorithm was used to solve for the optimal measurement light. 

While not the focus of their optimization, they did examine their recovery error - as we will do later -  in terms of errors in the CIELAB color space \cite{hunt2011measuring}. For the SFU reflectance set \cite{barnard2002data},  they found that their optimization method did not help them to significantly reduce $\Delta E^{*}_{ab}$ error (compared to using a non-optimized light). As we will report later, our optimization - based on a different mathematical formalism - does lead to significantly lower error for this data set.

\section{Matched Illumination} \label{sec:method}

Returning to Equation (\ref{eq:filtered_pixel}), it is apparent that we can think of a filter as modulating the spectral sensitivities of the sensors  - $F(\lambda)Q_k(\lambda)$ - or equivalently as modulating the spectral power of the light, $E(\lambda)F(\lambda)$. 
We call the modulated light, $E^m(\lambda)$ the {\it matched} illumination:
\begin{equation}
E^m(\lambda)=E(\lambda)F(\lambda).
\label{eq:matched}
\end{equation}
A camera with a filter $F(\lambda)$ placed in its optical pathway viewing the scene lit by a light $E(\lambda)$ makes the same measurement as the same camera without any filter but where the scene is illuminated by $E^m(\lambda)$ (assuming a simple viewing environment where we can ignore effects such as interreflection).

Let us move our development of the matched illumination idea to the discrete domain.
Given a $31\times 1$ illumination $\mathbf{e}$, we are looking for a matched illuminant $\mathbf{e}^m$ that makes the camera more colorimetric (more able to measure XYZs under the illuminant $\mathbf{e}$). Noting that
\begin{equation}
 \mathbf{e}^m=diag(\mathbf{e})\mathbf{f}.
\end{equation}
 
\noindent
Our optimization statement for the design of matched illuminations is written as:
\begin{equation}
    \argmin \limits_{   \mathbf{e}^m,\mathbf{M}}\parallel{ diag(\mathbf{e}^m)\mathbf{QM} -  diag(\mathbf{e})\mathbf{X}}\parallel^2_F
    \label{eq:opt_original}
\end{equation}
where $\parallel{}\parallel^2_F$ denotes the square of the Frobenius norm and, as before, $\mathbf{M}$ is a $3 \times 3$ full rank matrix.

\subsection{Simple Matched Illumination }

It is convenient to think of the lights (in a spectral illuminator) as a simple linear basis which can be used to describe a range of lights:
\begin{equation}
    \mathbf{e} = \mathbf{Bc} \;,\;\; \; 0\preceq \mathbf{c} \preceq 1 .
    \label{eq:light_linear_basis}
\end{equation}
For an illuminator with $k$ LED lights, $\mathbf{B}$ is a $31 \times k$ matrix. The $i$th column of the basis matrix $\mathbf{B}$ lists the maximum power of the $i$th LED light spectrum. $\mathbf{c}$ is a $k \times 1$ vector giving the intensity weights of the LED light channels.   
Additionally, of course, each coefficient is restrained by $c_i \in [0,1]$: it has to be between 0 and 100\% maximum power. In the simple basis world, we ignore the issue that the peaks of the basic light spectra shift as their intensity is changed. 

For a viewing illuminant $\mathbf{e}=\mathbf{B}\mathbf{c}$,  we can solve for the matched illumination $\mathbf{e}^m=\mathbf{B}\mathbf{c}^m$ (again $0\preceq \mathbf{c}^m \preceq 1$) by modifying Equation (\ref{eq:opt_original}): 
\begin{equation}
    \argmin \limits_{{\mathbf{c}^m}, \; {\mathbf{M}}}\parallel{ diag(\mathbf{B} {\mathbf{c}^m})\mathbf{Q}{\mathbf{M}} -  diag(\mathbf{e})\mathbf{X}}\parallel^2_F \;\;\; \text{s.t.}\;\; 0\preceq \mathbf{c}^m \preceq 1
    \label{eq:opt_illumination}
\end{equation}

To solve this optimization, we must estimate two unknown variables: the  coefficient vector $\mathbf{c}^m$ defining the matched illuminant and the $3 \times 3$ correction matrix $\mathbf{M}$. There is no closed-form solution to the problem. Analogously, to the prior art \cite{TIP2020filter}, we solve for $\mathbf{c}^m$ and $M$ using alternating least-squares regression: 

\begin{algorithm}[!ht]
	\caption{Algorithm for determining the channel weights in the simple model}
	\begin{algorithmic}[1]
		\State{$\mathbf{c}^m = \mathbf{c}^{guess}$, $\mathbf{M}={\cal I}_{3\times 3}$} \label{algo_init}
		\Repeat
	\State{$\mathbf{M}^{previous}= \mathbf{M}$ $\;,\;$ $\mathbf{c}^{previous}=\mathbf{c}^m$ }
	\State{$\min\limits_{\mathbf{M}} \parallel{diag(\mathbf{B} \mathbf{c}^{m}) \mathbf{QM} -  diag(\mathbf{e})\mathbf{X}}\parallel_{F}^2$}
		\label{code_M}	
     \State{$\min\limits_{\mathbf{c}^m} \parallel{diag(\mathbf{B} \mathbf{c}^m)\mathbf{Q M} -  diag(\mathbf{e})\mathbf{X}}\parallel_{F}^2,\;  0\preceq \mathbf{c}^m \preceq 1$} \label{codef}
  	\label{code_coef}
	\Until{$\parallel{diag(\mathbf{B} \mathbf{c}^{m}) \mathbf{QM} - diag(\mathbf{B} \mathbf{c}^{previous}) \mathbf{QM}^{previous}}\parallel_{F}^2 \, < \, \epsilon $}   \\
	 \Return{$\mathbf{c}^{m}$}
	\end{algorithmic}
	\label{algo1}
\end{algorithm}
First, we make an initial guess for the light coefficients (for the matched illumination). Then, it is straightforward to calculate the correction matrix $\mathbf{M}$ simply using the least-squares regression. Then we hold $\mathbf{M}$ fixed and solve for the optimal solution for $\mathbf{c}^m$ using Quadratic Programming \cite{luenberger1984linear} (to enforce the boundedness constraints).  The iteration continues until the difference between the current and previous solutions is below a criterion amount. The optimization is guaranteed to terminate.

A priori, we know that the peak lights - for some of the LEDs in our illuminator - do shift. But, if the shifts are small (generally they are), we should be able to adopt the simple algorithm and still obtain a good matched illuminant.

\subsection{Complex Matched Illumination}

In the complex model, we can still use the {\it basic} framework in Algorithm~\ref{algo1} to determine the matched illumination and the mapping matrix. However, we will address the problem that peak of the LED spectra shift as they are driven at different intensities.

To deal with the problem that the LED spectra shift, we will  measure the spectra power distribution emitted from each LED at a variety of intensity levels. Together, these spectra form an extended basis function set that better characterizes the illuminator system. We choose to use 10 uniform steps from 0 to maximum: $\mathbf{w} = [0, 0.1, 0.2, \cdots, 1]$.   With these measurements in hand, and given an arbitrary intensity level, we can use interpolation to estimate the  light spectrum -  for an arbitrary intensity level - as a convex combination of the two neighbouring intensities. For example, if we would like to know the spectrum at the intensity of 0.65, we calculate $\mathbf{e}_{0.65} = 0.5 * \mathbf{e}_{0.6} + 0.5* \mathbf{e}_{0.7}$ (where respectively $\mathbf{e}_{0.6}$ and $\mathbf{e}_{0.7}$ denotes an LED light driven at, respectively, 60 and 70\% of its maximum intensity)

Let us group all the measured lights into an array $\mathbf{A}$ with size of $31 \times 10 \times 11$, respectively $\#SampledWavelengths \times \#Channels \times  \#IntensityLevels$. We can extract a `local' basis from $\mathbf{A}$. For example, the $31\times 10$ maximum intensity basis (used in Algorithm \ref{algo1}) $\mathbf{B}=\mathbf{A}(:,:,11)$ where the `:' means to use all indices in that dimension (for those that use Matlab, we take the notation from there). The vectors $\mathbf{A}(:,5,7)$ and $\mathbf{A}(:,5,8)$ denote the 5th 
LED spectrum driven to 60\% and 70\% of the maximum intensities. Let us now define a normalized array of lights $\mathbf{A}^n$ where each light is divided by its intensity. As an example, $\mathbf{A}^n(:,5,8)=\mathbf{A}(:,5,8)/w_8$, which implies $w_8\mathbf{A}^n(:,5,8)=\mathbf{A}(:,5,8)$.

We use Algorithm \ref{algo2} to calculate the matched illuminant (when using an illuminator where the peaks change as function of intensity).
At initialization, we use a basis driven at their maximum intensities $\mathbf{B}=\mathbf{A}(:,:,11)$.  
 As in Algorithm 1, we calculate $\mathbf{M}$ and then
 we calculate the weight vector $\mathbf{c}^m$ - for the matched illumination - again using Quadratic Programming.  As the algorithm proceeds, we update the basis matrix $\mathbf{B}$.
 
\begin{algorithm}[!ht]
	\caption{Algorithm for determining the channel weights in the complex model}
	\begin{algorithmic}[1]
	\State{$\mathbf{B} = \mathbf{A}(:,11,:),\mathbf{c}^m = \mathbf{c}^{guess}$, $\mathbf{M}={\cal I}_{3\times 3}$}
	\Repeat
    \State{$\mathbf{M}^{previous}= \mathbf{M}$ $\;,\;$ $\mathbf{c}^{previous}=\mathbf{c}^m$ }
    \State{$\min\limits_{\mathbf{M}} \parallel{diag(\mathbf{B} \mathbf{c}^{m}) \mathbf{QM} -  diag(\mathbf{e})\mathbf{X}}\parallel_{F}^2$}
    \State{$\min\limits_{\mathbf{c}^m} \parallel{diag(\mathbf{B}\mathbf{c}^m)\mathbf{Q M}-  diag(\mathbf{e})\mathbf{X}}\parallel_{F}^2,\;  0\preceq \mathbf{c}^m \preceq 1$}  \label{algo2:update_coefs}
    \For{$i \leftarrow 1:\#Channels$}
    \If{$\mathbf{c}^{m}_{i} \leq \mathbf{w}_1$}
    \State{$\mathbf{B}(:,i) = \mathbf{A}^n(:,i,1)$}
     \Else{$ \; \; \mathbf{w}_{j-1} \leq \mathbf{c}^{m}_{i} < \mathbf{w}_j$}
       \State{$a = (\mathbf{c}^{m}_i - \mathbf{w}_{j-1})/(\mathbf{w}_{j}- \mathbf{w}_{j-1})$}
       \State{$\mathbf{B}(:,i) = (1-a)*\mathbf{A}^n(:,i,j-1) + a*\mathbf{A}^n(:,i,j)$}
       \EndIf
  \EndFor
  \Until{$\parallel{diag(\mathbf{B} \mathbf{c}^{m}) \mathbf{QM} - diag(\mathbf{B} \mathbf{c}^{previous}) \mathbf{QM}^{previous}}\parallel_{F}^2 \, < \, \epsilon _{2}$} \\
  \Return{$\mathbf{c}^{m}$}
  	\label{algo2_coef0}
	\end{algorithmic}
	\label{algo2}
\end{algorithm}

Depending on the coefficient vector value solved in Step \ref{algo2:update_coefs}, we have an indication of the basis that we `should' use. For example, if the $\mathbf{c}^m_6=0.5$, then this is proposing the 6th light (which on the first iteration has max power) should be driven at 50\% of the maximum intensity. Since we are aware of the spectral shift as the power changes, it makes sense to substitute the 50\% power spectrum (for the 6th light) into $\mathbf{B}$ for the next iteration. 
Actually, we substitute the power normalized spectrum $\mathbf{A}^n(:,6,5)$ for $\mathbf{B}(:,6)$. This is because on the next iteration if $\mathbf{c}^m_6 = 0.5$ we do not want to swap the basis again.
A similar algorithm was used by Mackiewicz and \emph{et al.} \cite{mackiewicz2012} to generate a light metamer for vision research. Interested readers are referred to the work for more details.

\begin{figure}[th]
    \centering
    \begin{subfigure}{0.48\textwidth}
    \centering
    \includegraphics[width=\textwidth]{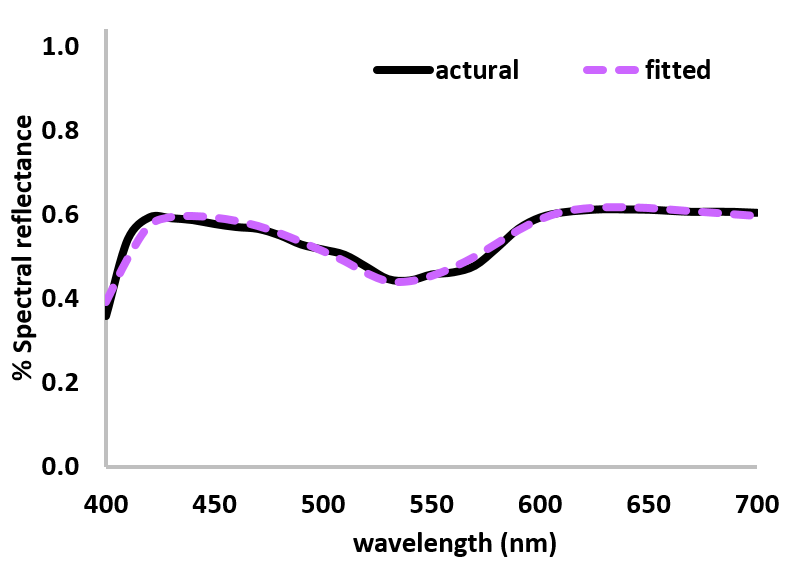}
    \caption{median error fitted reflectance (2.4\%)}
        \label{fig:SFU_median}
    \end{subfigure} 
    \begin{subfigure}{0.48\textwidth}
    \centering
    \includegraphics[width=\textwidth]{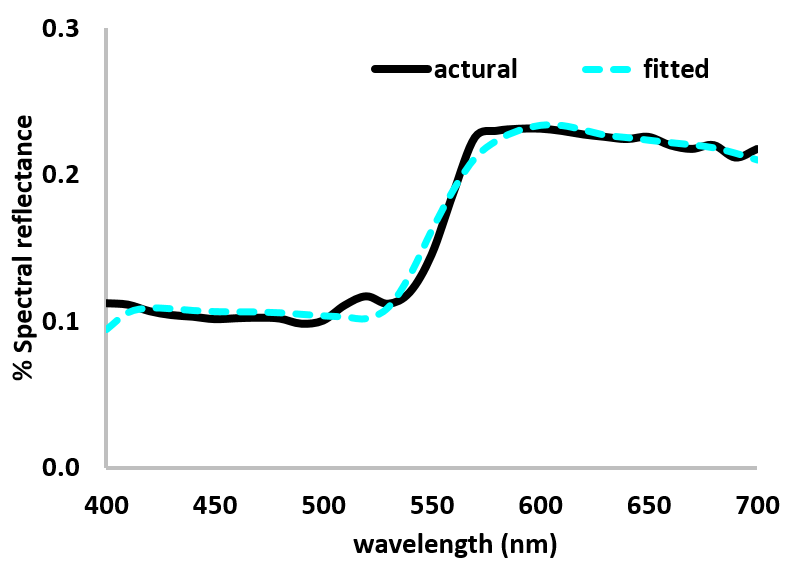}
    \caption{75-percentile error fitted reflectance (4.1\%)}
    \label{fig:SFU_75p}
    \end{subfigure} 
    \begin{subfigure}{0.48\textwidth}
    \centering
    \includegraphics[width=\textwidth]{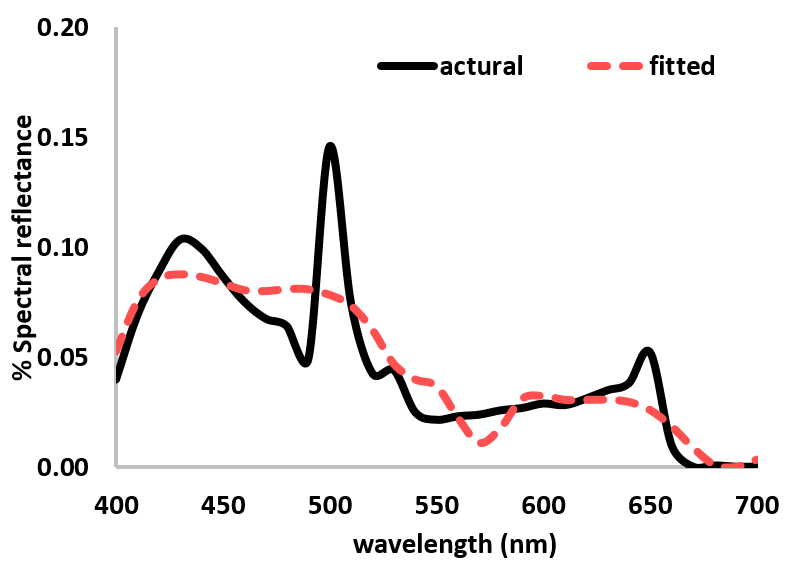}
    \caption{max error fitted reflectance (29\%)}
    \label{fig:SFU_max}
    \end{subfigure}
     \begin{subfigure}{0.48\textwidth}
    \centering
    \includegraphics[width=\textwidth]{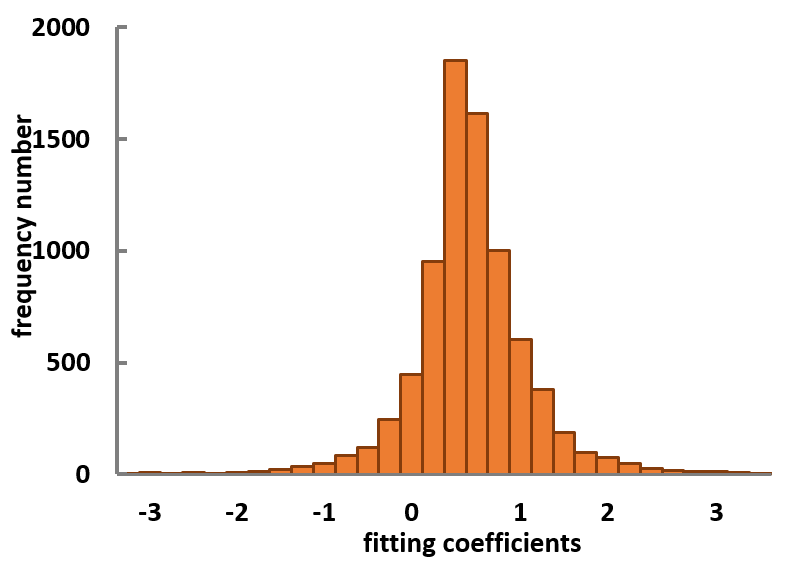}
    \caption{histogram of fitting coefficients}
    \label{fig:SFU_hist}
    \end{subfigure}
    \caption{In Panels (a), (b) and (c), the solid and dashed lines show respectively the actual and fitted reflectances for the median, 75-percentile and max error fits. Panel (d) shows the histogram of the fitting coefficients}
    \label{fig:SFU_reconstructed}
\end{figure}

\subsection{Algorithm for Making New Reflectance Data} \label{sec:synthesis}

In the next section, we will present synthetic and {\it real} color correction results for two object data sets: the \textbf{Macbeth} ColorChecker Chart \cite{mccamy1976chart} and 1995 reflectance spectra (\textbf{SFU1995}) \cite{barnard2002data}. The Macbeth chart is a standard chart used for characterizing and evaluating cameras. And \textbf{SFU1995} is a composite set comprising  1269 Munsell chips \cite{parkkinen1989}, 120 Dupont paint chips \cite{refl_vrhel}, 170 natural objects \cite{refl_vrhel}, 350 surfaces in \cite{krinov1947}, 24 Macbeth chart patches and 57 surfaces measured in Simon Fraser University.

The Macbeth Color checker only has 24 patches. And we do not - nor does anyone else - have access to the physical samples in the \textbf{SFU1995} data set. But, this reflectance set is often used to benchmark algorithms; so we'd like to quote real experimental results for \textbf{SFU1995}.

To bridge this experimental gap, we propose to describe the reflectances in \textbf{SFU1995} by a linear combination of no more than 4 color samples in the Macbeth data set:
\begin{equation}
\mathbf{r}_{target} \approx c_1 \mathbf{r}_1 + c_2 \mathbf{r}_2 + c_3 \mathbf{r}_3 + c_4 \mathbf{r}_4 = \mathbf{r}_{fit}
   \label{eq:linear_spectrum}
\end{equation}
where $\mathbf{r}_1, \mathbf{r}_2,\mathbf{r}_3$ and $\mathbf{r}_4$ are 4 spectra selected from the Macbeth chart and $\mathbf{r}_{target}$ is one of the reflectances in \textbf{SFU1995} . To simplify matters further, we used only 1 (out of 6) achromatic scale. Thus we wished to describe each reflectance in \textbf{SFU1995} as  a combination  of 4 (selected out of  19) Macbeth reflectances.

Assuming raw image capture, the RGB response to the target color can be calculated as the linearly composed RGBs of the 4 chosen color patches:
\begin{equation}
   \boldsymbol{\rho}_{target} \approx c_1 \boldsymbol{\rho}_1 + c_2 \boldsymbol{\rho}_2 + c_3 \boldsymbol{\rho}_3 + c_4 \boldsymbol{\rho}_4 .
   \label{eq:linear_RGBs}
\end{equation}
It follows that we can simulate the response to an unseen reflectance by applying the same linear combination - that approximates $\mathbf{r}_{target}$ - to the measured Macbeth RGBs. 

As a design choice, we choose to limit the number of reflectances to 4 in order to try and prevent linear combinations with large negative and positive coefficients (these coefficients could result in the RGB estimates to be susceptible to noise). 

In Figure \ref{fig:SFU_reconstructed}, we show three statistically representative reflectances (solid black lines) drawn from \textbf{SFU1995}  and three reconstructions (dashed lines) according to the method above. In order from panels (a) to (c), we are showing the median spectral error fit, the 75-percentile and the max error fit (where  the percentage error is defined as $\frac{||\mathbf{r}_{target}-\mathbf{r}_{fit}||}{||\mathbf{r}_{target}||}$). In panel (d), we show the histogram of the coefficients for the 1995 reflectances (where each reflectance is fit with 4 different Macbeth spectra). The coefficients are in the range [-3,3] indicating that any noise increase in the transformed RGBs will be small. Notice the peak of the histogram seems to be at 0. Actually,  there are not many coefficients that are exactly zero,  the histogram bin counts a range of coefficients. Evidently, some Macbeth reflectances make only small contributions to the linear combination matching to a target reflectance. 

Overall, the fit of Macbeth reflectances to the \textbf{SFU1995} is surprisingly good. 

\begin{figure}[tb]
    \centering
    \begin{subfigure}{0.48\textwidth}
    \centering
    \includegraphics[width=\textwidth]{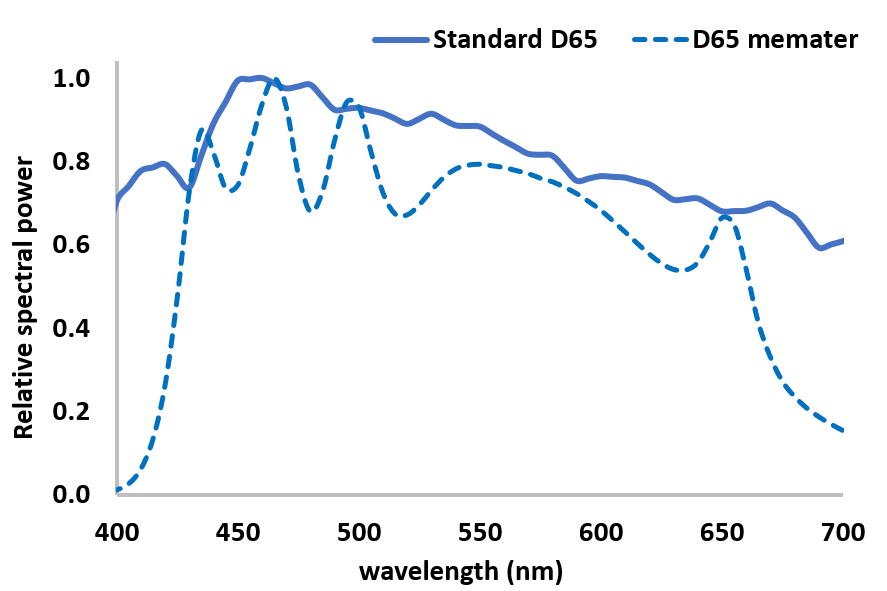}
    \caption{D65 and its metamer}
        \label{fig:d65_metamer}
    \end{subfigure} 
    \begin{subfigure}{0.48\textwidth}
    \centering
    \includegraphics[width=\textwidth]{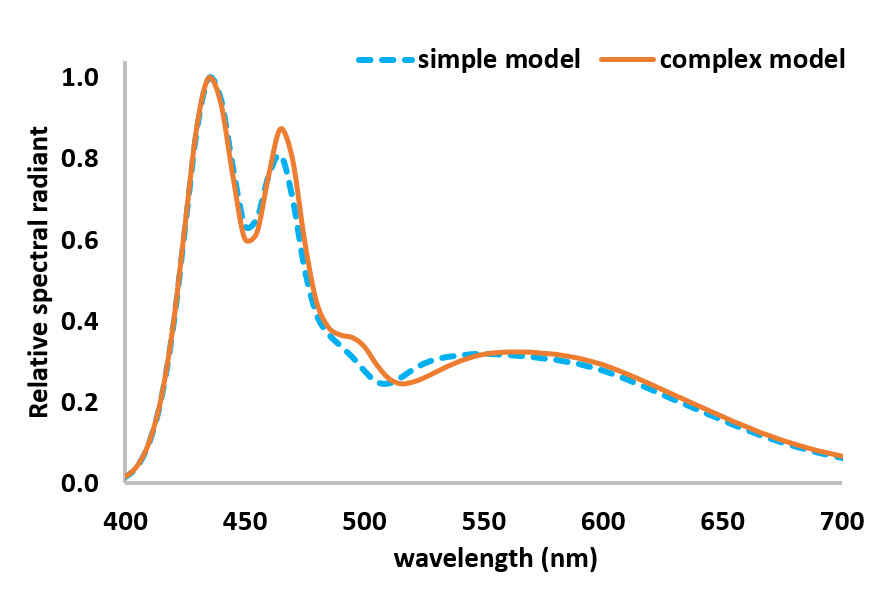}
    \caption{matched illuminations}
    \label{fig:matchedIllum}
    \end{subfigure} 
    \caption{(a) Relative spectral power distributions of the CIE D65 illuminant (solid line) and its metamer (dashed line) generated by the LED illuminator. (b) The \emph{matched} illuminations solved by the simple (dashed line) and complex models (solid line).}
    \label{fig:fig4}
\end{figure}

\section{Results} \label{sec:exp}

A D65 illuminant metamer generated by the LED Illuminator system is shown in Figure~\ref{fig:d65_metamer} (where its maximum power is normalized to one). In the figure, we also plot the theoretical CIE D65 in solid line. A D65 metamer is a spectrum that produces the same XYZ tristimulus values (relative XYZ tristimulus values of [0.9385,	1.0000,	1.0472]) as the theoretical D65 illuminant; yet, they are different in the spectral composition \cite{wyszecki1982color}. 
Our illuminator can make many D65 metamers \cite{metamer2014}, here we chose the metamer that has the least spectral error from the standard D65. 

In Figure~\ref{fig:matchedIllum}, we show the matched illuminations calculated with respect to   the simple  and complex illuminator models (respectively, solved using Algorithms \ref{algo1} and \ref{algo2}). From the figure, we see that the two matched illuminants are similar. There are, however, small spectral differences in the range of 450\,nm to 550\,nm.  Both matched lights are even bluer (have more radiant power in the short-wave  region of the visible spectrum) than the actual measurement D65 metamer.

\begin{figure}[th]
    \centering
    \begin{subfigure}{0.48\textwidth}
    \centering
    \includegraphics[width=\textwidth]{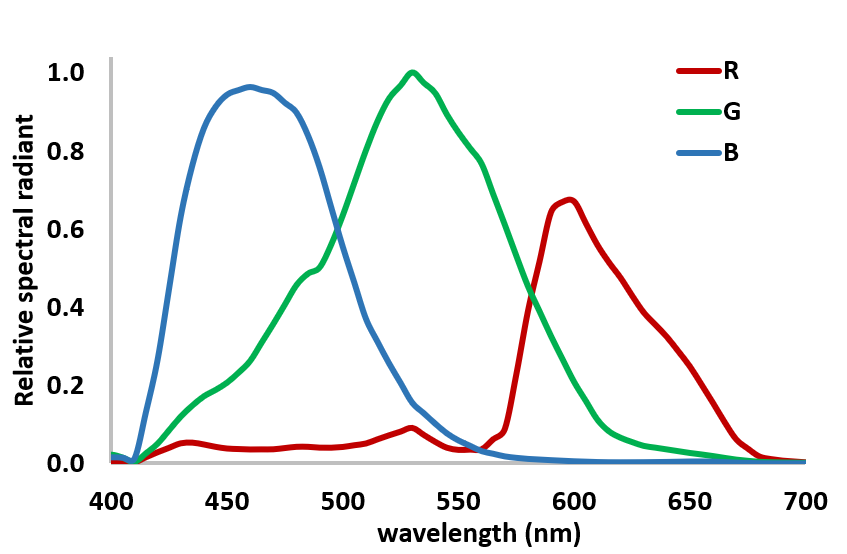}
    \caption{camera sensitivity functions}
        \label{fig:NikonD5100}
    \end{subfigure} 
    \begin{subfigure}{0.48\textwidth}
    \centering
    \includegraphics[width=\textwidth]{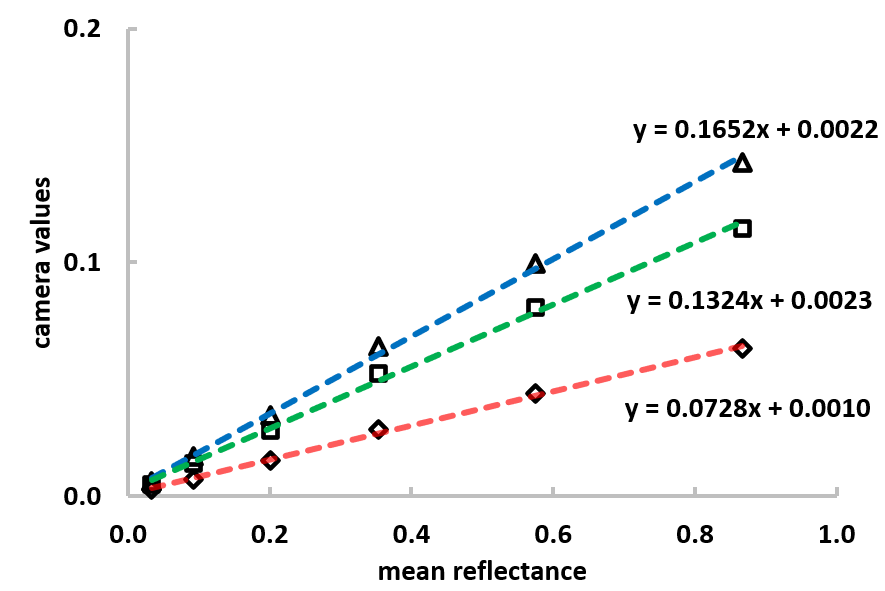}
    \caption{linearity of camera responses}
        \label{fig:camera_linearity}
    \end{subfigure} 
    \caption{(a) The spectral sensitivity functions of a Nikon D5100 DSLR camera. (b) The linearity property of the camera responses with respect to six neutral colors on the Macbeth color chart captured under the D65 illuminant.}
    \label{fig:camera}
\end{figure}

\subsection{Simulated Experiments} \label{sec:simulated_exp}

\begin{table}[bth]
\centering
\caption{$\Delta E_{ab}^{*}$  statistics  of simulated color correction performance for two testing datasets when using the color corrected  native camera under D65 metamer, the color corrected camera with the matched illuminations generated by the illuminator system under the simple and complex models for a Nikon D5100 DSLR camera.}
\renewcommand{\arraystretch}{1.25}
\setlength\tabcolsep{2pt}
\begin{tabular}{lcccc|cccccc}
\hline
 & \multicolumn{3}{c}{Macbeth chart}  & &  &\multicolumn{5}{c}{SFU1995 surfaces} \\ \hline
 & Mean & Median &Max & & & Mean & Median & 95\% & 99\%  & Max   \\ \hline
D65 Metamer   & 1.54	&1.62	&4.88 & &  & 1.61 & 0.92   & 5.23 & 11.70 & 21.28 \\ \hline
Simple matched illuminant      & 0.86	&0.58	&4.11 & &
& 0.97 & 0.61   & 3.03 & 5.02  & 7.80  \\  \hline
Complex matched illuminant     &0.80	&0.57	&3.71 & &
    & 0.86 & 0.49   & 2.73 & 5.04  & 17.87  \\ \hline
\end{tabular}
\label{tab:DE_simulated}
\end{table}

In the simulated experiments, we evaluated how well  a  Nikon camera (see Figure \ref{fig:NikonD5100}) can measure the colors of  two object sets: the {\bf Macbeth} ColorChecker Chart and 1995 reflectance spectra (\textbf{SFU1995}). The reflectance data  of the {\bf Macbeth} chart were measured by a Konica Minolta spectrophotometer CM700d  in the range of 400\,nm to 700\,nm for every 10\,nm at our laboratory. And \textbf{SFU1995} surface data were collated at Simon Fraser University \cite{barnard2002data}.

We first calculated the camera RGB responses of the Nikon camera according to Equation (\ref{eq:pixelformation1}) provided the spectral data of the D65 metamer, the matched illuminations, the reflectances, and the camera. In the tables that follow we respectively call the matched illuminations derived using Algorithms 1 and 2 the {\it Simple} and {\it Complex} matched illuminations. Also, the corresponding ground-truth XYZ values under the D65 illuminant metamer were computed.

Our  three sets of camera RGBs - for the D65 metamer and the simple and complex matched illuminants - were separately mapped (color corrected) using least-squares regression to estimate the  XYZs. The predicted and the ground-truth XYZs were converted into the CIELAB color space and then the color difference between them was evaluated in terms of $\Delta E_{ab}^{*}$ \cite{ohta2006colorimetry}. The error statistics were calculated over all test reflectances.

The results of this experiment  are summarized in Table \ref{tab:DE_simulated}. The left and right of the table report the experiments for, respectively, the \textbf{Macbeth} and  \textbf{SFU1995} data sets. We calculated the CIELAB  $\Delta E_{ab}^{*}$ errors for three cases. First, when the native camera RGBs - recorded under the D65 metamer - were color corrected to XYZs. Then we color corrected the RGBs measured under the Simple and Complex matched illuminations. The Mean, Median and Max errors are shown for both reflectance sets. For the \textbf{SFU1995} set which has a much larger number of reflectances, we also calculated the 95- and 99-percentile errors.

From Table \ref{tab:DE_simulated}, it is clear that measuring and then color correcting RGBs measured under the  \emph{matched illumination}   lead to better color measurement accuracy compared to the original D65 metamer.
For the {\bf Macbeth} color chart, respectively by the simple and complex matched illuminations, we find there is a reduction of  44\% and 48\% in terms of mean $\Delta E_{ab}^*$  error, 64\% and 65\% for median $\Delta E_{ab}^*$  error, and 24\% and 16\% for max $\Delta E_{ab}^*$  error.  For the {\bf SFU1995} dataset, the 95- and 99- percentile errors are substantially improved; they are halved for the complex matched illumination. 

Overall, there is a modest improvement when the complex (as opposed to the simple) matched illumination is used. This is encouraging - and serendipitous - since it is the complex model that actually corresponds to the physical properties of the illuminator we have in our lab.

\subsection{Experiments Using  Measured Data} \label{sec:real_exp}

Images of the {\bf Macbeth} checker - under the D65 Metamer and the simple and complex matched illuminants - were captured with a Nikon D5100 DSLR camera. The camera used a  fixed focal length of 35 mm with  f-number of 5, ISO at 1600 and exposure time at 1/40 s. 
To check camera linearity, we  plot mean reflectance, for the 6 achromatic colors on the {\bf Macbeth} chart - against the mean of their RGB responses, see Figure \ref{fig:camera_linearity}. The dashed lines are the linear fitting curves with its function shown in the figure. We see that the curves almost pass through the origin which confirms the good linearity of our camera.

To obtain the RGBs that we will use in our experiments,  raw  Nikon image files (NEF) of the Macbeth chart were captured,  converted and demosaiced into TIFF format using DCRAW \cite{dcraw}. Then the camera raw RGBs of a selected area of about $200 \times 200$ pixels were averaged for each color patch in the Macbeth chart. 
To ensure lighting uniformity, we also captured images of an X-rite White Balance chart placed at the same spot as the Macbeth chart. By dividing out the RGBs in the checker by the corresponding RGBs measured in the  white chart, we corrected for non-uniform illumination. Of course dividing by white can be thought of as multiplying by a diagonal matrix (whose diagonal components are the reciprocal of the RGBs in the white reference chart). This, however, does not change our color correction optimization. If $\mathbf{M}$ denotes a $3\times 3$ matrix optimally mapping the RGBs of a camera under a given light to the corresponding XYZs and we then multiply the RGBs by a diagonal matrix $D$, then least-squares color correction will return $\mathbf{D}^{-1}\mathbf{M}$. That is, the output from color correction will be the same.

For the ground-truth values, we used the same  XYZs that were calculated for the synthetic experiments (discussed in Section \ref{sec:simulated_exp}). Our ground truth are the XYZ tristimuli of the Macbeth color checker illuminated by the D65 metamer.

We now repeat the color correction experiment for real RGB data. But, we used the  method set forth in Section \ref{sec:synthesis} to allow us to investigate the performance for the {\bf SFU1995} dataset. That is, we model each SFU reflectance as a linear sum of four Macbeth reflectances. Because of the linearity of capture, applying the same linear combination to  the corresponding Macbeth RGBs will result in an RGB that corresponds to the linearly combined reflectances. In this way, given the measurements from a Macbeth checker, we can test the matched illuminant approach on a much larger reflectance dataset. Also, although computing RGBs in this way will increase noise, we are averaging the responses over 4$200 \times 200$ pixels; so the effect of the noise is negligible.

In Table \ref{tab:DE_experimental}, we see that the error between the color corrected native RGBs and the the ground truth XYZs is significantly higher compared to those in the simulated experiments. This is to be expected. We are choosing our matched illuminant based on estimated spectral sensitivities and measured illuminants and there will certainly be discrepancies in both. Further, although care is taken to measure the color checker to minimize any specular reflectance, there is likely a small specular component in our data (not present in the synthetic experiment).

\begin{table}[tbh]
\centering
\caption{Error statistics $\Delta E_{ab}^{*}$  of experimental measured data for two object sets for the color corrected  native camera under \textbf{ D65 Metamer}, the color corrected camera with the matched illuminations generated by the illuminator system under the simple and complex models for a Nikon D5100 DSLR camera.}
\renewcommand{\arraystretch}{1.25}
\setlength\tabcolsep{2pt}
\begin{tabular}{lcccc|cccccc}
\hline
 & \multicolumn{3}{c}{Macbeth chart}  & &  &\multicolumn{5}{c}{reconstructed SFU1995 surfaces} \\ \hline
 & Mean & Median &Max & & & Mean & Median & 95\% & 99\%  & Max   \\ \hline
D65 Metamer   
&2.38	&1.84	&5.66 & &  &2.62	&1.76	&7.64	&16.08	&40.15 \\ \hline
Simple matched illuminant      & 1.93	 &1.53	&4.36 & &
&1.99	&1.39	&5.49	&9.44	&45.33  \\ \hline
Complex matched illuminant     &1.85	&1.67	&4.08 & &
    &1.82	&1.22	&4.84	&9.37	&55.85 \\ \hline
\end{tabular}
\label{tab:DE_experimental}
\end{table}

\begin{figure}[th]
    \centering
    \begin{subfigure}{0.48\textwidth}
    \centering
    \includegraphics[width=\textwidth]{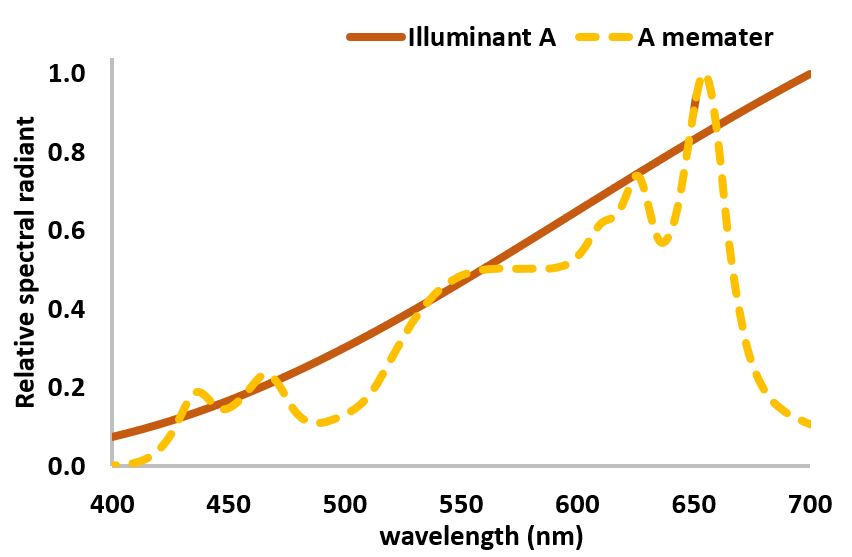}
    \caption{A and its metamer}
        \label{fig:A_metamer}
    \end{subfigure} 
	\hfill
\begin{subfigure}{0.48\textwidth}
    \centering
    \includegraphics[width=\textwidth]{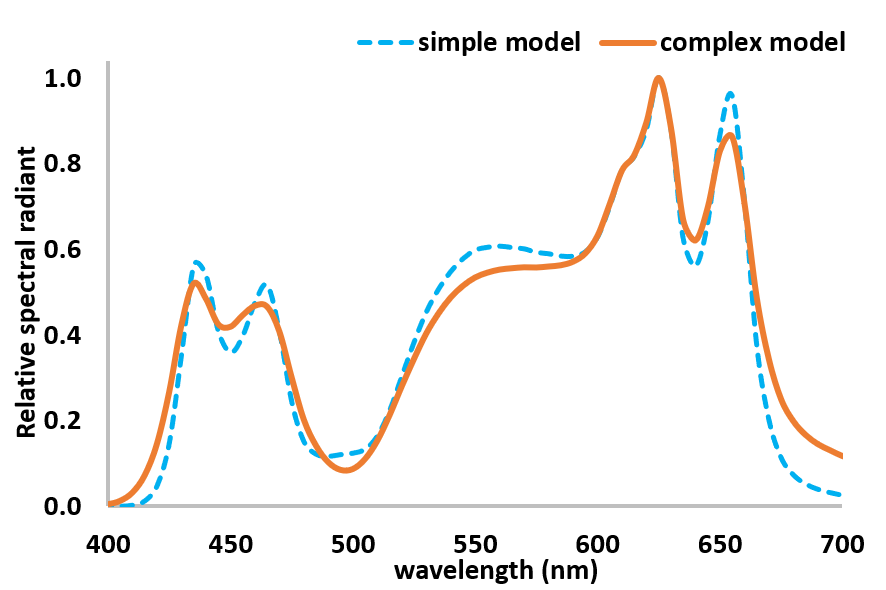}
    \caption{matched illuminations}
    \label{fig:matchedIllumA}
    \end{subfigure} 
    \caption{(a) The relative spectral power distribution of the CIE Illuminant A (solid line) and its metamer (dashed line) generated by the LED illuminator. (b) The \emph{matched} illuminations solved by the simple (dashed line) and complex models (solid line).}
    \label{fig:illumA}
\end{figure}

Significantly, when we measure and correct real RGBs measured under matched illuminations, we record significantly lower $\Delta E_{ab}^{*}$ color errors.   
The error for the {\bf Macbeth} reflectances is reduced by a modest amount (e.g. 22\% for the mean metric by the complex matched illuminant). The performance difference for the larger {\bf SFU1995} reflectance set is larger: the corrected RGBs (measured under the D65 Metamer) are, for the mean, median and 99-percentile errors, respectively 44\%, 44\% and 72\% higher compared with the measurements taken under the complex matched illumination. As before, we find the complex matched illumination condition leads to the lowest errors overall.

\subsection{Color Correction under CIE Illuminant A}

We repeated the synthetic and real experiments of Section \ref{sec:simulated_exp} and \ref{sec:real_exp} for a CIE Illuminant A. Again, we calculated the metamer for CIE A and solved for the simple and complex matched illuminants, see Figure \ref{fig:illumA}.

Our experimental results are summarized in Tables \ref{tab:DE_simulated_A} and \ref{tab:DE_experiment_A}. We see the trend of the data is the same as those of Illuminant D65. As before, there is a significant improvement in color measurement error when colors are recorded and corrected under matched illuminants. And, once more, we find there is a small advantage of using the complex illuminator model.

\begin{table}[bth]
\centering
\caption{$\Delta E_{ab}^{*}$  statistics  of simulated color correction performance for two testing data sets when using the color corrected  native camera under \textbf{A Metamer}, the color corrected camera with the matched illuminations generated by the illuminator system under the simple and complex models for Nikon D5100 DSLR camera.}
\renewcommand{\arraystretch}{1.25}
\setlength\tabcolsep{2pt}
\begin{tabular}{lcccc|cccccc}
\hline
 & \multicolumn{3}{c}{Macbeth chart}  & &  &\multicolumn{5}{c}{SFU1995 surfaces} \\ \hline
 & Mean & Median &Max & & & Mean & Median & 95\% & 99\%  & Max   \\ \hline
A Metamer   & 1.55	&1.07	&5.91 & &  & 1.51 & 0.79   & 4.68 & 13.17 & 21.59 \\ \hline
Simple matched illuminant      & 0.65	&0.48	&2.35 & &
& 0.61 & 0.38   & 2.04 & 4.61  & 7.31  \\  \hline
Complex matched illuminant     &0.63	&0.50	&2.25 & &
    & 0.58 & 0.36   & 1.97 & 4.04  & 6.08  \\ \hline
\end{tabular}
\label{tab:DE_simulated_A}
\end{table}

\begin{table}[tbh]
\centering
\caption{Error statistics $\Delta E_{ab}^{*}$  of experimental tests for two object data sets for the color corrected  native camera under A metamer, the color corrected camera with the matched illuminations generated by the illuminator system under the simple and complex models for Nikon D5100 DSLR camera.}
\renewcommand{\arraystretch}{1.25}
\setlength\tabcolsep{2pt}
\begin{tabular}{lcccc|cccccc}
\hline
 & \multicolumn{3}{c}{Macbeth chart}  & &  &\multicolumn{5}{c}{reconstructed SFU1995 surfaces} \\ \hline
 & Mean & Median &Max & & & Mean & Median & 95\% & 99\%  & Max   \\ \hline
A Metamer   
&2.61	&2.43	&6.35 & &  &2.60	&1.73	&7.42	&17.48	&34.79 \\ \hline
Simple matched illuminant      & 1.55	 &1.25	&3.94 & &
&1.95	&1.19	&5.78	&11.73	&55.08  \\ \hline
Complex matched illuminant     &1.45	&1.09	&3.90 & &
&1.87	&1.15	&5.57	&10.93	&49.74 \\ \hline
\end{tabular}
\label{tab:DE_experiment_A}
\end{table}

\section{Conclusion} \label{sec:conclusion}

In prior work (e.g. \cite{zhuCIC2018,TIP2020filter}), it has been shown that it is possible to design a color prefilter that when it is  placed in the optical path of a camera it can make the camera {\it almost} colorimetric. However, none of the filters previously designed have been manufactured. And, it is not known to what extent they can be manufactured.

In this paper, we pose the filter-design problem in an equivalent form. We propose that placing a filter in front of a light source is broadly equivalent to placing the filter in front of the camera. Since we now have tunable multi-spectral LED illuminators, we can model the function of the filter as a modulation of the light source. For a given measurement light and a camera, we show how we can optimally modulate a light source to solve for a {\it Matched Illumination}. The matched illumination for D65 is spectrally quite different but results in RGBs which are more able to be color corrected to CIE XYZ tristimuli than RGBs measured under D65.

Experiments validate our results. On synthetic and real data, we show that there is a significant benefit (up to 50\%) in using a matched illumination for measuring color (under a desired measurement light). A novel aspect of our experimental methodology is that we show how the measurements made for a Macbeth ColorChecker chart can be used to calculate results for a much larger reflectance dataset.

\bibliography{mybib}

\begin{thebibliography}{10}
\newcommand{\enquote}[1]{``#1''}

\bibitem{ives1915}
H.~E. Ives, \enquote{The transformation of color-mixture equations from one
  system to another,} {\protect\JournalTitle{Journal of the Franklin
  Institute}} \textbf{180}, 673--701 (1915).

\bibitem{luther1927}
R.~Luther, \enquote{Aus dem gebiet der farbreizmetrik,}
  {\protect\JournalTitle{Zeitschrift Technische Physik}} \textbf{8}, 540--558
  (1927).

\bibitem{HORN1984}
B.~K.~P. Horn, \enquote{Exact reproduction of colored images,}
  {\protect\JournalTitle{Computer Vision, Graphics, and Image Processing}}
  \textbf{26}, 135 -- 167 (1984).

\bibitem{ohta2006colorimetry}
N.~Ohta and A.~Robertson, \emph{Colorimetry: fundamentals and applications}
  (John Wiley \& Sons, 2006).

\bibitem{nakamura2016image}
J.~Nakamura, \emph{Image sensors and signal processing for digital still
  cameras} (CRC press, 2016).

\bibitem{farrell1995method}
J.~E. Farrell and B.~A. Wandell, \enquote{Method and apparatus for identifying
  the color of an image,}  (1995). {} U.S. Patent 5479524.

\bibitem{wu2000imaging}
W.~Wu, J.~P. Allebach, and M.~Analoui, \enquote{Imaging colorimetry using a
  digital camera,} {\protect\JournalTitle{Journal of Imaging Science and
  Technology}} \textbf{44}, 267--279 (2000).

\bibitem{macadam1945colorimetric}
D.~L. MacAdam, \enquote{Colorimetric specifications of {W}ratten light
  filters,} {\protect\JournalTitle{Journal of the Optical Society of America
  A}} \textbf{35}, 670--675 (1945).

\bibitem{hardeberg2004filter}
J.~Y. Hardeberg, \enquote{Filter selection for multispectral color image
  acquisition,} {\protect\JournalTitle{Journal of Imaging Science and
  Technology}} \textbf{48}, 105--110 (2004).

\bibitem{Imai2001DigitalCF}
F.~H. Imai, S.~Quan, M.~R. Rosen, and R.~S. Berns, \enquote{Digital camera
  filter design for colorimetric and spectral accuracy,} in \emph{University of
  Joensuu,}  (2001), pp. 13--16.

\bibitem{vora1997design}
P.~L. Vora and H.~J. Trussell, \enquote{Mathematical methods for the design of
  color scanning filters,} {\protect\JournalTitle{IEEE Transactions on Image
  Processing}} \textbf{6}, 312--320 (1997).

\bibitem{xu2016filter}
P.~Xu and H.~Xu, \enquote{Filter selection based on light source for
  multispectral imaging,} {\protect\JournalTitle{Optical Engineering}}
  \textbf{55}, 074102 (2016).

\bibitem{martinez2019spectral}
M.~{\'A}. Mart{\'\i}nez-Domingo, M.~Melgosa, K.~Okajima, V.~J. Medina, and
  F.~J. Collado-Montero, \enquote{Spectral image processing for museum lighting
  using {CIE LED} illuminants,} {\protect\JournalTitle{Sensors}} \textbf{19},
  5400 (2019).

\bibitem{TIP2020filter}
G.~D. Finlayson and Y.~Zhu, \enquote{Designing color filters that make cameras
  more colorimetric,} {\protect\JournalTitle{IEEE Transactions on Image
  Processing}} \textbf{30}, 853--867 (2021).

\bibitem{filter2020sensor}
Y.~Zhu and G.~D. Finlayson, \enquote{A mathematical investigation into the
  design of prefilters that make cameras more colorimetric,}
  {\protect\JournalTitle{Sensors}} \textbf{20}, 6882 (2020).

\bibitem{Wang2021}
L.~Wang, A.~Sole, J.~Y. Hardeberg, and X.~Wan, \enquote{Optimized light source
  spectral power distribution for {RGB} camera based spectral reflectance
  recovery,} {\protect\JournalTitle{Optics Express}} \textbf{29}, 24695--24713
  (2021).

\bibitem{barnard2002data}
K.~Barnard, L.~Martin, B.~Funt, and A.~Coath, \enquote{A data set for color
  research,} {\protect\JournalTitle{Color Research \& Application}}
  \textbf{27}, 147--151 (2002).

\bibitem{mccamy1976chart}
C.~S. McCamy, H.~Marcus, and J.~G. Davidson, \enquote{A color-rendition chart,}
  {\protect\JournalTitle{Journal of Applied Photographic Engineering}}
  \textbf{2}, 95--99 (1976).

\bibitem{hong2001study}
G.~Hong, M.~R. Luo, and P.~A. Rhodes, \enquote{A study of digital camera
  colorimetric characterization based on polynomial modeling,}
  {\protect\JournalTitle{Color Research \& Application}} \textbf{26}, 76--84
  (2001).

\bibitem{finlayson2015color}
G.~D. Finlayson, M.~Mackiewicz, and A.~Hurlbert, \enquote{Color correction
  using root-polynomial regression,} {\protect\JournalTitle{IEEE Transactions
  on Image Processing}} \textbf{24}, 1460--1470 (2015).

\bibitem{hung1993colorimetric}
P.-C. Hung, \enquote{Colorimetric calibration in electronic imaging devices
  using a look-up-table model and interpolations,}
  {\protect\JournalTitle{Journal of Electronic Imaging}} \textbf{2}, 53--62
  (1993).

\bibitem{andersen2016weighted}
C.~F. Andersen and D.~Connah, \enquote{Weighted constrained hue-plane
  preserving camera characterization,} {\protect\JournalTitle{IEEE Transactions
  on Image Processing}} \textbf{25}, 4329--4339 (2016).

\bibitem{drew1992natural}
M.~S. Drew and B.~V. Funt, \enquote{Natural metamers,}
  {\protect\JournalTitle{CVGIP: Image Understanding}} \textbf{56}, 139--151
  (1992).

\bibitem{hunt2011measuring}
R.~W.~G. Hunt and M.~R. Pointer, \emph{Measuring Colour} (John Wiley \& Sons,
  2011), 4th ed.

\bibitem{luenberger1984linear}
D.~G. Luenberger and Y.~Ye, \emph{Linear and nonlinear programming} (Springer,
  2015), 4th ed.

\bibitem{mackiewicz2012}
M.~Mackiewicz, S.~Crichton, S.~Newsome, R.~Gazerro, G.~D. Finlayson, and
  A.~Hurlbert, \enquote{Spectrally tunable {LED} illuminator for vision
  research,} in \emph{Conference on Colour in Graphics, Imaging, and Vision,}
  (Society for Imaging Science and Technology, 2012), pp. 372--377.

\bibitem{parkkinen1989}
J.~P.~S. Parkkinen, J.~Hallikainen, and T.~Jaaskelainen,
  \enquote{Characteristic spectra of {M}unsell colors,}
  {\protect\JournalTitle{Journal of the Optical Society of America A}}
  \textbf{6}, 318--322 (1989).

\bibitem{refl_vrhel}
M.~J. Vrhel, R.~Gershon, and L.~S. Iwan, \enquote{Measurement and analysis of
  object reflectance spectra,} {\protect\JournalTitle{Color Research \&
  Application}} \textbf{19}, 4--9 (1994).

\bibitem{krinov1947}
E.~L. Krinov, \enquote{Spectral reflectance properties of natural formations,}
  Tech. rep., National Research Council of Canada (1947).

\bibitem{wyszecki1982color}
G.~Wyszecki and W.~S. Stiles, \emph{Color science: concepts and methods,
  quantitative data and formulae} (Wiley New York, 1982), 2nd ed.

\bibitem{metamer2014}
G.~Finlayson, M.~Mackiewicz, A.~Hurlbert, B.~Pearce, and S.~Crichton,
  \enquote{On calculating metamer sets for spectrally tunable {LED}
  illuminators,} {\protect\JournalTitle{Journal of the Optical Society of
  America A}} \textbf{31}, 1577--1587 (2014).

\bibitem{dcraw}
D.~Coffin, \emph{Decoding raw digital photos in Linux}. Available from
  https://www.dechifro.org/dcraw/.

\bibitem{zhuCIC2018}
G.~D. Finlayson, Y.~Zhu, and H.~Gong, \enquote{Using a simple colour pre-filter
  to make cameras more colorimetric,} in \emph{Color and Imaging Conference,}
  (Society for Imaging Science and Technology, 2018), pp. 182--186.

\end{thebibliography}

\end{document}